\documentclass[prc,aps,nofootinbib,showkeys,showpacs,twocolumn]{revtex4}   
\usepackage{epsfig}  
\usepackage{graphicx}  
\usepackage{amssymb}
\usepackage{color}  
\begin{document}  
  
\title{Effect of pairing on one- and two-nucleon transfer below the Coulomb barrier: a time-dependent microscopic description.}

\author{Guillaume Scamps}  
 \email{scamps@ganil.fr}  
\affiliation{GANIL, CEA/DSM and CNRS/IN2P3, Bo\^ite Postale 55027, 14076 Caen Cedex, France}  
\author{Denis Lacroix} \email{lacroix@ganil.fr}  
\affiliation{GANIL, CEA/DSM and CNRS/IN2P3, Bo\^ite Postale 55027, 14076 Caen Cedex, France}  
    
\begin{abstract}  
The effect of pairing correlation on transfer reaction below the Coulomb barrier is investigated qualitatively and 
quantitatively using a simplified version of the Time-Dependent Hartree-Fock + BCS approach. The effect of particle number 
symmetry breaking on the description of reaction and dedicated methods to extract one and two-nucleon transfer probabilities ($P_{1n}$
 and  $P_{2n}$) 
 in a particle number symmetry breaking approach are discussed. Influence of pairing is systematically investigated in the $^{40}$Ca+ $^{40,42,44,46,48,50}$Ca reactions.
A strong enhancement of the two-particle transfer probabilities due to initial pairing correlations is observed. This enhancement 
induces an increase of the ratio of probabilities $P_{2n} / (P_{1n})^2$ compared to the case with no pairing. 
It is shown that this ratio increases strongly 
as the center of mass energy decreases with a value that could be larger than ten in the deep sub-barrier regime. 
An analysis of the pair transfer sensitivity to the type of pairing interaction, namely 
surface, mixed or volume, used 
in the theory is made. It is found that the pair transfer is globally insensitive to the type of force and mainly depends 
on the pairing interaction strength. 
\end{abstract}

  
\maketitle  
  
\section{Introduction}  

The possibility to access cross section much below the Coulomb barrier 
has revealed new aspects like the hindrance of fusion cross section (see for instance \cite{Jia06}) whose origin is 
still debated \cite{Mis06, Ich07}. Among possible interpretation, other competing processes like single- or multi-nucleon transfer 
might eventually be enhanced and/or modify the capture process \cite{Sag12,Sag12a}. 
New experimental observations \cite{Oer01,Lem09,Cor11} in the moderate and deep sub-barrier regime might lead to important 
new insight especially on the process of pair transfer. The description of such pair transfer is particularly complex since it requires to treat 
the quantum tunneling of a composite, eventually correlated, system. In particular, pairing correlations among last bound nucleons is anticipated to 
play a crucial role. Following the pioneering work of Refs \cite{Bes66,Bro68,Rip69,Bro73}, an important effort is currently being made to improve the 
description of pair transfer in superfluid systems \cite{Dob96,Kha04,Pot09,Kha09,Ave08,Mou11,Shi11,Pot11,Pll11,Gra12,Gam12}. These approaches 
have usually in common that transition probabilities from the initial to the final nucleus are estimated using state of the art Hartree-Fock Bogolyubov (HFB) 
and Quasi-Particle Random Phase Approximation (QRPA) 
nuclear models while the reaction dynamics part is treated in a completely separated steps using coupled channels technique. 

The present work is an attempt to treat nuclear structure and nuclear reaction aspects in a common microscopic 
framework that includes pairing.   
Recently, active research has been devoted to include pairing correlations 
into the nuclear dynamics using the Time-Dependent HFB (TDHFB) approach \cite{Has07,Ave08,Ste11}. While current
applications can be performed in an unrestricted space, due to 
the required effort, applications of TDHFB have been essentially made on process involving 
one nucleus, like giant resonances.  The use of TDHFB to nuclear reactions remains tedious.
A simplified version of TDHFB based on the BCS approximation is considered. 
This theory has been proposed already some times ago \cite{Blo76}  and recently 
applied with some success either to collective motion in nuclei \cite{Eba10}, to reactions in 
1D models  \cite{Sca12}. First step toward collisions have been reported in ref. \cite{Eba12}. The TDHF+BCS approach has the advantage to be simpler than 
the original TDHFB theory while keeping part of the physics of pairing. Note that, time-dependent microscopic theories have several 
advantages compared to other techniques. Many effects, like possible dynamical deformation or core polarization 
during the reaction are automatically accounted for. In addition, other competing phenomena 
like emission to the continuum and/or fusion are simultaneously treated.  Since, many aspects of the theory applied here 
have been extensively discussed in Refs. \cite{Eba10, Sca12}, only main aspects features are recalled below. 

\section{Nuclear reactions with pairing}

Time-Dependent Hartree-Fock (TDHF) has become a standard tool 
to describe nuclear reactions like fusion or transfer reactions 
(see \cite{Sim10a} and ref. therein). In the present work, the {\rm TDHF3D} code 
of ref. \cite{Kim97} is extended to include pairing correlations.  Below, specific aspects 
related to the introduction of pairing are discussed.  

\subsection{Initial conditions}
\label{sec:static}

The reaction is simulated on a 3-dimensional mesh. Following 
the standard procedure \cite{Kim97}, the two nuclei are initiated 
separately and then positioned consistently with the desired impact parameter $b$
and center of mass energy $E_{\rm c.m.}$.  The initial wave function can be written as 
\begin{eqnarray}
| \Psi(t_0) \rangle &=& | \Phi_1(t_0) \rangle \otimes | \Phi_2(t_0) \rangle 
\end{eqnarray} 
where $ | \Phi_\alpha (t_0) \rangle$ denotes the many-body wave-function of nucleus $\alpha=1,2$. Usually, these wave-functions corresponds to Slater 
determinants. It is assumed here to take the more general form of a quasi-particle vacuum written as
\begin{eqnarray}  
|\Phi_\alpha (t_0) \rangle =\prod_{k>0} \left( u^\alpha_k(t_0) + v^\alpha_k(t_0) a^\dagger_k(t_0) a^\dagger_{\bar  
k} (t_0)\right)|-\rangle.  
\end{eqnarray}  
where $a^\dagger_k(t_0)$ stands for the creation operator associated to the canonical single-particle states, denoted 
hereafter by $| \varphi_k(t_0) \rangle$ 
while $(u_k(t_0), v_k(t_0))$ are the standard upper and lower components of the quasi-particle states. Note that due to the spatial separation of 
the two nuclei, a common single-particle basis can be used. Accordingly, we can omit the $\alpha$ index and directly write the total wave-function as:
\begin{eqnarray}
| \Psi(t_0) \rangle &=&\prod_{k>0} \left( u_k(t_0) + v_k(t_0) a^\dagger_k(t_0) a^\dagger_{\bar  k} (t_0)\right)|-\rangle.  \label{eq:fullwf}
\end{eqnarray} 

In practice, initial states for each nucleus have been obtained using the {\rm EV8} code \cite{Bon05} that solve the self-consistent BCS 
equations in the Energy Density Functional framework \cite{Ben03}. 
Single-particle states are written in $\mathbf r$-space and spin space, denoted by 
$\sigma = \uparrow, \downarrow$ as:
\begin{eqnarray}
a^\dagger_k &=& \sum_\sigma \int d{\bf r} \varphi_k({\bf r}, \sigma)\Psi^\dagger_\sigma({\bf r}),
\end{eqnarray}
where $\Psi^\dagger_\sigma({\bf r})$ are standard spinors creation operators. In {\rm EV8}, time-reversal symmetry is assumed 
and single-particle states can be grouped by pairs of time-reversed states $(k, \bar k)$. 
Associated quasi-particle creation operators $(\beta^\dagger_k, \beta^\dagger_{\bar k})$
are written using the following convention for the Bogolyubov transformation:
\begin{eqnarray}
\left\{
\begin{array} {cc}
\beta^\dagger_{k}  = & \sum_{{\bf r}} u_k ( {\bf r} , t_0 )  \Psi^\dagger_{\uparrow} ({\bf r}) + v_k ({\bf r},t_0)  \Psi_{\downarrow} ({\bf r}), \\
\\
\beta^\dagger_{\bar k}  = & \sum_{{\bf r}} u_k ( {\bf r} , t_0 )  \Psi^\dagger_{\downarrow} ({\bf r}) - v_k ({\bf r},t_0)  \Psi_{\uparrow} ({\bf r}), \\
\end{array}
\right.
\end{eqnarray} 
where, using time-reversal properties,  we have 
\begin{eqnarray}
u_k ( {\bf r} , t_0) & = & u_k \varphi_{\bar k} ( {\bf r} ,\uparrow)   =  u_k \varphi_k ( {\bf r} ,\downarrow) , \\
v_k ( {\bf r}  , t_0) & = &  v_k  \varphi^*_{k} ( {\bf r} ,\downarrow)  = v_k \varphi^*_{\bar k} ( {\bf r} ,\uparrow). 
\end{eqnarray}

\begin{figure}[htbp] 
	\centering\includegraphics[width= \linewidth]{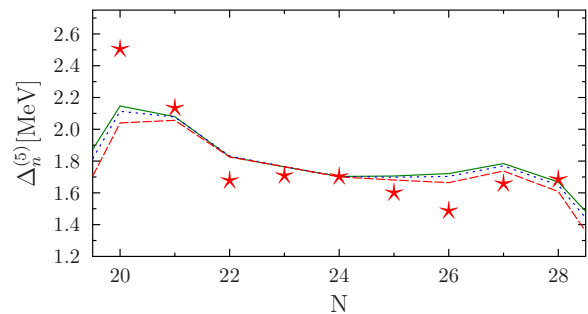}
  	\caption{Experimental (red star) and theoretical neutron gaps for $Z=20$ as a function of $N$. 
	The volume, mixed and surface interaction are respectively shown by solid, dotted and dashed lines. }
	\label{fig:gapCa} 
\end{figure}

The Skyrme Sly4d functional \cite{Kim97} is used in the mean-field 
channel while for pairing, the following effective neutron-neutron interaction is used:
\begin{eqnarray}
V_\tau ({\mathbf r}, \sigma; {\mathbf r'} ,\sigma') &=& V^{\tau \tau}_0 \left(1 - \eta \frac{\rho([{\mathbf r}+{\mathbf r'}]/2)}{\rho_0} \right)\delta_{{\mathbf r},{\mathbf r'}}
\left[1-P_{\sigma
\sigma'} \right] \nonumber
\end{eqnarray}   
where $P_{\sigma \sigma'}$ is the spin exchange operator and where $\rho_0 = 0.16$ fm$^{-3}$. Here $\tau=n,p$ stands for neutron or proton channel, only neutron-neutron and proton-proton
interaction are considered.
Three different forces, standardly called volume ($\eta=0$), mixed ($\eta=0.5$) and surface ($\eta=1$) 
will be used below. In each case, the neutron pairing interaction strength $V^{nn}_0$ was adjusted to properly reproduce the experimental gap for the calcium 
isotopic chain deduced from masses using the $5$ points formula \cite{Dug01}. 
Theoretical odd systems binding energies have been computed using blocking techniques.   
Values of the interaction parameters are reported in table \ref{tab1:trans}.
The proton interaction strength is taken from ref. \cite{Ber09} but do not play 
any role due to the proton closed shell. 
\begin{table}[!h]
\begin{center}
\begin{tabular}{|c|c|c|c|}
 \hline
     interaction & $\eta$ & $V_0^{nn}$ [MeV.fm$^3$] &  $V_0^{pp}$ [MeV.fm$^3$] \\
 \hline
 \hline
    volume & 0 & 585  & 490 \\
 \hline
    mixed & 0.5 & 798  & 755 \\
 \hline
    surface & 1 & 1256  & 1462 \\
 \hline
\end{tabular}
\end{center} 
\caption{Parameters of the neutron-neutron and proton-proton pairing strength used 
in the present work.}
\label{tab1:trans}
\end{table}
Illustration of the pairing gap obtained for the three results of the fit are shown in Fig. \ref{fig:gapCa}  for the three types of pairing interaction. 
A comparison of the neutron pairing gap $\Delta^{(5)}_n$  obtained  using the different interactions is done in Fig. \ref{fig:gapCa}.
The three interactions lead to gaps that are compatible with each others and compatible with the experimental gaps along
the calcium isotopic chains especially in the mid-shell.  In the following, we consider systematically reactions between the doubly 
magic $^{40}$Ca nucleus and other $^{4x}$Ca isotopes.  The two reactions $^{40}$Ca+$^{40}$Ca and $^{40}$Ca+$^{48}$Ca
will correspond to reactions between two normal systems, while in other cases, one of the nucleus will present 
pairing.  

\subsection{Time-dependent equation of motion}

Once the two nuclei have been initiated, the reaction is simulated by performing the dynamical
evolution of the many-body wave-packet given by Eq. (\ref{eq:fullwf}). Here,  the TDHF+BCS approximation that may be derived from a   
variational principle \cite{Blo76} or by an approximate  
reduction of the TDHFB equations \cite{Eba10} is used. Since, properties as well as numerical aspects of
the TDHF+BCS method are discussed in Refs. \cite{Eba10,Sca12}, only main ingredients of the 
theory are summarized here. In this theory, the wave-function remains at all time in its canonical basis, Eq. (\ref{eq:fullwf}), and the single-particle 
states evolution identifies with the mean-field dynamics with:
\begin{eqnarray}  
i\hbar \partial_t | \varphi_k \rangle = (h[\rho]-\eta_k) | \varphi_k \rangle  
\label{eq:wf_ev}
\end{eqnarray}  
 where $\eta_k(t) = \langle \varphi_k(t) |h[\rho]|  
\varphi_k(t) \rangle$ is a time-dependent phase that is conveniently chosen to minimize the effect of the 
$U(1)$ symmetry breaking. $h[\rho]$ corresponds here to the self-consistent mean-field derived from the Skyrme 
functional including time-odd components.

Along the dynamical path, the information is contained in the normal and anomalous densities, denoted by $\rho$
and $\kappa$ written in ${\mathbf r}$-space as: 
 \begin{eqnarray}
\rho_{\sigma\sigma'}({\mathbf r}, {\mathbf r'}) & = & \sum_{k \gtrless 0 } n_k \varphi^*_{k}({\mathbf r}, \sigma) \varphi_{k}({\mathbf r'}, \sigma')  \\
\kappa_{\sigma\sigma'}({\mathbf r}, {\mathbf r'}) &=& \sum_{k > 0} \kappa_k ( \varphi_{k}({\mathbf r}, \sigma) \varphi_{\bar k}({\mathbf r'}, \sigma') \nonumber \\
&& - \varphi_{\bar k}({\mathbf r}, \sigma) \varphi_{ k}({\mathbf r'}, \sigma') ).
\end{eqnarray}
$(k,\bar k)$ corresponds to pair of single-particle states that were originally degenerated in the static calculation due to 
time-reversal symmetry. $n_k = v^2_k$ denote the occupation numbers while $\kappa_k = u_k^* v_k$ are the components 
of the anomalous density in the canonical basis. Conjointly to the single-particle evolution, the equation of motion of the components $(u_k,v_k)$
or equivalently of $(n_k, \kappa_k)$ should be specified. Following Ref. \cite{Eba10}, we have:
\begin{eqnarray}  
i \hbar \frac{d}{dt} n_k(t) &=&   
\kappa_{k}(t) \Delta_{k}^*(t) - \kappa_{k}^*(t) \Delta_{k} (t), \\  
i \hbar \frac{d}{dt} \kappa_k(t) &=&   
\kappa_k(t) ( \eta_k(t) + \eta_{\overline{k}}(t) ) + \Delta_{k}(t) (2n_k(t)-1) ,\label{eq:k-bcs}  \nonumber
\end{eqnarray} 
where $\Delta_k(t)$ correspond to the pairing field components given by 
\begin{eqnarray}
\Delta_k(t) & = & - \sum_{l>0} \overline{v}_{k\overline{k}l\overline{l}} \kappa_k(t) g_k(t_0).
\end{eqnarray}
$g_k$ corresponds to the cut-off function that select the pairing window. This cut-off should be taken consistently with 
the static calculation \cite{Bon05}.  Here, a slightly different prescription is used compared to the original {\rm EV8} 
with 
\begin{eqnarray}
g_k(t_0) = f(\eta_k(t_0) - \lambda) f( \lambda - \eta_k(t_0)) \theta(-\eta_k(t_0)) .
\end{eqnarray}
$f$ here corresponds to a Fermi distribution with a cutoff  at $5$ MeV and a stiffness parameter equal to $0.5$ MeV \cite{Bon05}, while 
$\theta(\eta)$ equals one for $\eta > 0$ and zero elsewhere. This additional cut-off insures that only states that are initially bound 
are considered during the evolution.

As discussed in ref. \cite{Sca12}, the reduction of the TDHFB to TDHF+BCS leads to some inconsistencies, especially 
regarding the one-body continuity equation, making the interpretation of the dynamics difficult. To avoid this problem, 
we used here the Frozen Occupation Approximation (FOA). In the FOA, it is assumed that the main effect of 
pairing originates from the initial correlations that induce partial occupations of the orbitals and non-zero components of the 
two-body correlation matrix, denoted by $C_{12}$. 
Possible reorganization in time of occupation numbers and components of $C_{12}$ are neglected. Said differently, 
occupation numbers $n_k$ and components $\kappa_k$ are kept fixed in time and equal to their initial values. 
Note that similar ideas have been used recently to describe two-particle break-up reaction using the Time-Dependent 
Density-Matrix approach\cite{Ass09}.
This simplification 
is motivated by the fact that (i) it solves the problem of continuity equation \cite{Sca12} (ii) in the simple one dimensional 
model considered in the same reference, it gives rather good description of the emission of particles and is sometimes more predictive than
the full TDHFB theory (iii) the FOA approximation applied to collective motion in nuclei \cite{Sca12b} gives results that are very close to the full TDHF+BCS dynamics  
reported in \cite{Eba10}.      

 \begin{figure}[htbp] 
	\centering\includegraphics[width= \linewidth]{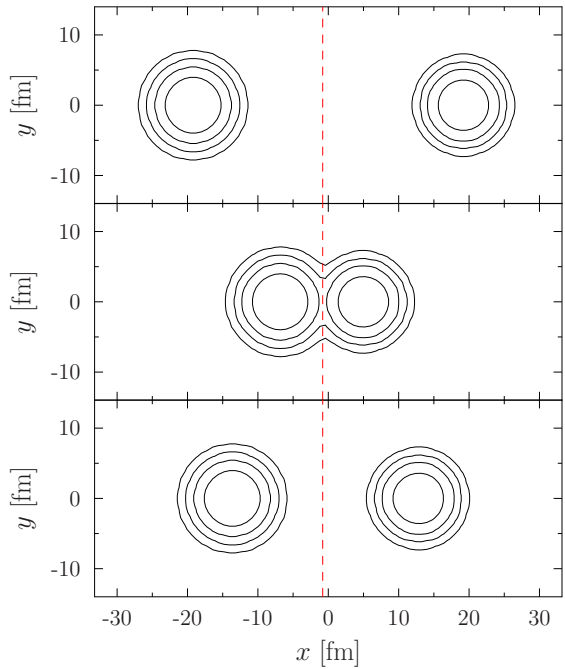}
  	\caption{Evolution of the neutron density projected onto 
	the reaction plan $z=0$, for the reaction $^{46}$Ca+$^{40}$Ca at impact parameter $b=0$ fm 
	and center of mass energy $E_{\rm c.m.} = 49$ MeV. 
	At initial time (top), $t=20\times 10^{-22}$s (middle) and $t=37\times 10^{-22}$s (bottom). 
	The neck position is indicated by the dashed vertical line.} 
	\label{fig:film} 
\end{figure}
\subsection{Illustration of reactions}

In the present work, we are interested in reactions below the Fusion barrier like the one presented in Ref. \cite{Cor11} where the probabilities to transfer 
$x$ neutrons, denoted by $P_{xn}$  can be extracted as a function of the minimal distance of approach $D$ during the collision. Assuming a Coulomb trajectory, $D$
is related to the center of mass energy $E_{\rm c.m.}$ through:
\begin{eqnarray}
D=\frac{Z_P Z_T e^2}{2 E_{\rm c.m.}}\left( 1 + \frac1{\sin(\theta_{cm}/2)} \right)
\end{eqnarray}
where $Z_P$ and $Z_T$ are the target and projectile proton number while $\theta_{cm}$ is the center of mass scattering angle.  Following 
Ref. \cite{Sim10b}, only central collisions will be considered here and different distances $D$ are simulated by varying 
the center of mass energy. 
Initial conditions are obtained on a lattice of $2L_x\times 2L_y \times 2L_z = 22.4\times22.4\times22.4$ fm$^3$ noting that the {\rm EV8} code uses symmetries to reduce the calculation in one octant of this space. 
The dynamical evolution are performed in the center of mass frame using a Runge-Kutta 4 algorithm 
on a spatial grid of $L_x\times L_y \times 2L_z =60.8\times22.4\times22.4$ fm$^3$ with a lattice spacing $\Delta x=0.8$ fm. The time-step is 
$\Delta t=0.015\times 10^{-22}$ s.
Note that, non-equilibrium particle emission is negligible due to the small center of mass energy in the entrance channel. 

As an illustration, the neutron density profiles of the reaction  $^{46}$Ca+$^{40}$Ca are shown at different stages of the reaction in Fig. \ref{fig:film}. 
During the reaction, the two nuclei approach from each other, 
stick together during a certain time and then re-separate. During the contact time that strongly depends on the initial center of mass energy, they eventually exchange particles. 

\subsection{Particle transfer probability in normal systems}
\label{sec:normal}

In practice, the system can be cut into two pieces at the neck position to calculate the expectation value of 
the number of exchanged nucleons 
from one-side to the other. By convention, we will denote by $B$ the subspace where the lightest nucleus is initially 
(right side of the neck position in Fig. \ref{fig:film}) and ${\bar B}$ 
the rest of the total space. In a mean-field approach, the simplest way to obtain the number of exchanged 
particles is to estimate the operator $\hat N_B$ defined through\cite{Das79}:
\begin{eqnarray} 
  \hat{N}_B=\sum_\sigma \int d{\bf r} \Psi^\dagger_\sigma({\bf r}) \Psi_\sigma({\bf r}) \Theta({\bf r})
\end{eqnarray} 
with the time-dependent 
wave-function (\ref{eq:fullwf}). Here $\Theta({\bf r})$ is zero on the left side of the neck and 1 elsewhere.

\begin{figure}[htbp] 
	\centering\includegraphics[width= \linewidth]{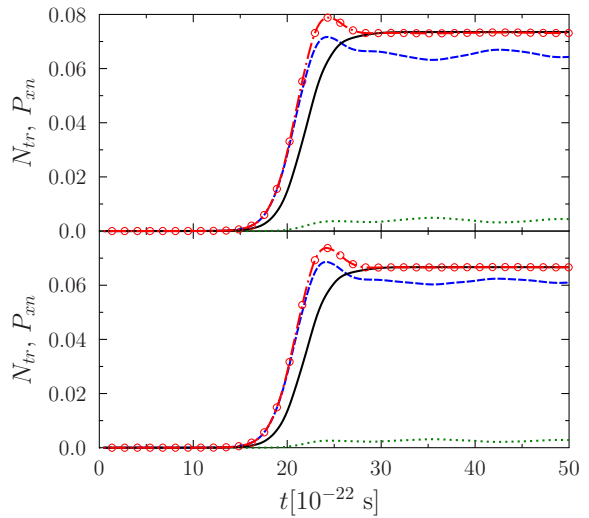}
  	\caption{Evolution of the mean number of particles transferred from $^{46}$Ca to $^{40}$Ca as a function of time 
	during the reaction illustrated in Fig. \ref{fig:film}.  The probability $P_{1n}$ (dashed line) and $P_{2n}$ (dotted line) to exchange 1 and 2 particles 
	obtained by making projection on the side $B$ are also presented as well as the quantity $P_{1n} + 2 P_{2n}$ (open circles).  Top: mean-number of 
	particles and probabilities obtained without projecting on good particle number in the total space. Bottom: same with an additional projection on neutron number 
	$N=46$ in the total space. } 
	\label{fig:trans46} 
\end{figure}

An illustration of the mean number of transferred particle, denoted by $N_{\rm tr} \equiv \langle  \hat{N}_B \rangle -20$, 
from $^{46}$Ca to $^{40}$Ca at center of mass energy $E_{\rm c.m.} = 49$ MeV is shown in Fig. \ref{fig:trans46} (solid line).
As discussed in ref. \cite{Sim10b}, a deeper understanding of the transfer process can be achieved by introducing 
projection onto good particle numbers in the sub-space $B$ (or equivalently $\bar B$).
The projection operator on a given number of particles $N$ inside the subspace $B$ can be written
as (see \cite{Sim10b}) 
\begin{eqnarray} 
  \hat{P}_B(N)= \frac{1}{2\pi}\int_{0}^{2\pi}d\varphi e^{i\varphi(\hat{N}_B-N)},
\end{eqnarray} 
where $\varphi$ is the standard gauge angle.
Then the probability to find $N$ particles in the subspace $B$ is:
\begin{eqnarray} 
P_B(N)&=& \langle \Psi(t) | \hat{P}_B(N) |  \Psi(t) \rangle \nonumber \\
&=& \frac{1}{2\pi}\int_{0}^{2\pi}d\varphi e^{-i\varphi N} \langle  \Psi(t)| \Psi_B(\varphi , t) \rangle
\end{eqnarray}
where $| \Psi_B(\varphi, t) \rangle=e^{i\varphi\hat{N}_B} | \Psi(t) \rangle$ is a new quasi-particle vacuum obtained from the original one 
by making a rotation $\varphi$ in the gauge space from the original state.

The probabilities extracted by projection are linked to the mean number of particles through the sum rule:
\begin{eqnarray}
\langle \hat N_B \rangle & = & \sum_N N P_B(N). \label{eq:sumrule}
\end{eqnarray}  
Usually, experimental data are presented in terms of probabilities to exchange $1$, $2$, ... $x$ neutrons (resp. protons), denoted respectively by $P_{1n}$,
$P_{2n}$, ... $P_{xn}$ (resp. $P_{1p}$, $P_{2p}$, ... $P_{xp}$). In the present reaction, these probabilities are defined through $P_{xn} = P_B (20+x)$ while 
the above sum rule reads $N_{\rm tr} = \sum_x x P_{xn}$.       

In the present work, probabilities have been evaluated using the Pfaffian technique of ref. \cite{Ber12} and explicit formulas for the wave-packet 
are given in appendix \ref{app:proba}.  An illustration of $P_{1n}$ and $P_{2n}$ probabilities obtained using the projection method is shown in top panel of Fig. \ref{fig:trans46} for the $^{46}$Ca. 
As already noted in ref. \cite{Sim10b}, the $1n$ and $2n$ channels are often dominating over other multi-nucleon transfer channels 
leading to $N_{\rm tr} \simeq P_{1n} + 2 P_{2n}$, that is perfectly fulfilled in Fig. \ref{fig:trans46} after the two nuclei re-separate.

\subsection{Particle transfer probability in superfluid systems}

\begin{figure}[htbp] 
	\centering\includegraphics[width=0.8\linewidth]{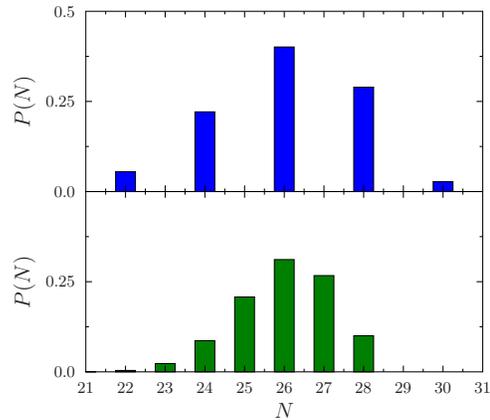}
  	\caption{Illustration of the distribution of probabilities to have $N$ particles initially for the $^{40}$Ca + $^{46}$Ca reaction in the 
TDHF+BCS case (top) and TDHF with initial filling approximation (bottom). Note that since $^{40}$Ca has a good particle number, $N$ is defined here 
as the number of particles in the wave-function describing the $^{46}$Ca and is centered around $N=26$. } 
	\label{fig:distN} 
\end{figure}

Strictly speaking, the above method to extract transfer probabilities is only valid for normal systems, i.e. when the 
wave-function (\ref{eq:fullwf}) identifies with a Slater determinant that is an eigenstate of particle number. For nuclei that present 
pairing, the initial wave-function explicitly breaks the particle number symmetry and the BCS states is obtained by imposing the particle number 
only in average. This is for instance the case for the $^{46}$Ca discussed above. Said differently, the ground state that is used for $^{46}$Ca 
not only presents a component with $N=26$ neutrons but also with surrounding number of neutrons. These components lead to spurious contributions 
in the probabilities extracted in previous section.  A possible way to remove this contamination is to first select the relevant component 
with $N_0=20+26$ particles in the full space and then consider the projection onto different particle numbers in the sub-space $B$. In the following, we denote 
by $\hat P(N_0)$ the projector on $N_0$ particles in the full space:
\begin{eqnarray} 
  \hat{P} (N_0)= \frac{1}{2\pi}\int_{0}^{2\pi}d\varphi e^{i\varphi(\hat{N}-N_0)},
\end{eqnarray} 
where $\hat N$ is now the complete particle number operator. More generally, to estimate the possible effect of contribution from
components $N \neq N_0$, one can compute the probability $P(N)$ that the initial state belongs to the Hilbert space of $N$ particles. This
probability is defined through:
\begin{eqnarray}
P(N) &=& \langle \Psi(t_0) | \hat{P} (N)| \Psi(t_0) \rangle 
\end{eqnarray}    
and is shown in Fig. \ref{fig:distN} (top panel). Since $^{40}$Ca has a well defined number of particles, by convention, $N$ in the x axis of Fig. \ref{fig:distN}
is taken here as the number of particles of its collision partner. 
Only even components are non-zero due to the specific form of
the state (Eq. (\ref{eq:fullwf})). While the distribution is properly centered around the imposed mean number of particles, non negligible contributions
coexist, especially for $N = N_0 \pm 2$ in the initial state.  

To remove possible influence of these spurious components, it is possible to define at all time a state with good number of neutrons 
\begin{eqnarray}
| N_0(t)\rangle \equiv \frac{1}{\sqrt{\langle \Psi (t) | \hat{P} (N_0)| \Psi(t)\rangle }}  \hat{P} (N_0)
| \Psi (t)\rangle.
\end{eqnarray}  
Then, the mean number of transferred particle as well as probabilities $P_{xn}$ can be computed using the same technique as in section \ref{sec:normal}.
Note that, the double projection approach proposed here can be regarded as a first step towards Projection After Variation  
(PAV) approach standardly applied in nuclear structure generalized here to binary reactions.
An illustration of result is given in bottom panel of Fig. \ref{fig:trans46}. The comparison of projected (bottom) and unprojected (top) panel, show that
$N_{\rm tr}$ and $P_{1n}$ are only slightly affected by the removal of spurious components. This is a  quite general feature 
we observed in applications presented in the article. However, the difference between $P_{2n}$ with and without projection can  
be as large as several orders of magnitude.
This conclusion also holds for a larger number of particles transferred. 

A second difficulty arises, that could already be seen in Fig. \ref{fig:trans46}, when pairing is non-zero. While $N_{\rm tr}$ 
after collisions converges to a well defined asymptotic value, small oscillations of $P_{1n}$ and $P_{2n}$ around their asymptotic 
values remain. These oscillations are also present if the expectation value $\langle \hat N^2_B \rangle$ is computed as a function of time with or without 
projection onto good particle number in the total space. This problem points out a difficulty in theories like TDHF+BCS. In a previous article 
\cite{Sca12}, we have shown that the one-body continuity equation is always respected in TDHFB, while in TDHF+BCS, it is respected only if single-particle 
occupations are frozen, which is the case in the present work. However, these theories provide only approximate treatment of the two-body density matrix and 
in particular do not respect the two-body continuity equation. This difficulty is not specific to the TDHF+BCS theory but is also present in 
TDHFB. Indeed, we have checked in the 1D model developed in ref. \cite{Sca12}, adapted to treat transfer, that similar oscillations occur 
even if the full TDHFB is solved. In the following, results obtained for nuclei with non-vanishing pairing will be presented with error-bars with height 
equal to oscillation amplitudes. In most cases displayed below, error-bars will be too small to be seen.

 \subsection{Sensitivity to the pairing residual interaction} 

\begin{figure}[htbp] 
	\centering\includegraphics[width= \linewidth]{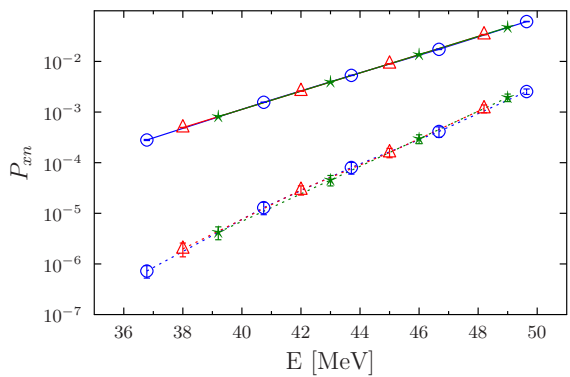}
  	\caption{Comparison of asymptotic probabilities $P_{1n}$ (solid line) and $P_{2n}$ (dotted line) obtained with the three pairing interactions 
	for the $^{40}$Ca + $^{46}$Ca at various center of mass energy below the fusion barrier: volume  (open circles), mixed (open triangles) and surface (crosses) interactions are represented by open circles, open triangles and stars respectively. Note that error-bars due to final time oscillations are also presented but are very small. 
	   } 
	\label{fig:inter} 
\end{figure}

Three different pairing interactions, presented in section \ref{sec:static} have been used to initialize the collision partners.
These interactions lead to different spatial properties of the pairing field but have been adjusted to reproduce the experimental gaps (see Fig. \ref{fig:gapCa}).
In figure \ref{fig:inter}, asymptotic values of one- and two-nucleon transfer probabilities are reported as a function of center of mass energy for the $^{40}$Ca + $^{46}$Ca 
for the three pairing interactions below the Fusion barrier.  
As seen in the figure, the extracted transfer probabilities are insensitive to the type of interaction used. It turns out, that whatever is
the form of the pairing effective zero-range vertex, if the interaction is carefully adjusted to reproduce the same experimental gap (Fig. \ref{fig:gapCa}),
the final transfer rate is also the same.  Note that the present finding is not in contradiction with ref. \cite{Gra12}  where different types of interactions 
(mixed and surface) were shown to give different two-particle transfer from ground state to ground state. The two forces used in ref.  \cite{Gra12} have been adjusted 
to reproduce the same two neutrons separation energies but lead to different pairing gap. In the present work, we do not see any evidence of a dependence of the pair transfer process on the shape of the pairing force that is used.

Since all types of force lead to the same probabilities, below only results of one of the interaction (mixed)
are shown.  

\subsection{The no-pairing limit}

Here, we are interested in the enhancement of pair transfer probabilities as the pairing is introduced 
in the transport theory. To quantify this enhancement, it is necessary to also perform calculation without 
pairing interaction, i.e. TDHF. An additional difficulty arises in the comparison between systems with and without pairing. 
Quite often, especially when a given j-shell is partially occupied, nuclei initialized with {\rm EV8} in the Hartree-Fock limit are 
deformed. The introduction of pairing stabilizes the spherical shape. Therefore, a direct comparison 
of the case with and without pairing not only probes the effect of 
pairing but also the effect of deformation that is (i) not correct for calcium isotopes (ii) 
not the objective of the present work.  

To avoid, possible effects of deformation, we used the filling approximation for the last occupied shell, i.e. we assume that 
the last shell has partial occupations $n_k$ such that all angular momentum projections $m$ are occupied in the same way. 
This insures the convergence of the mean-field theory towards non-deformed systems. This approach implies that 
the initial system is not anymore described by a wave-packet like in Eq. (\ref{eq:fullwf}), that would identify with a Slater determinant 
in the usual TDHF, but by a many-body density matrix of the form:
\begin{eqnarray}
\hat D(t) & = & \frac{1} {Z} \exp(-\sum_k \lambda_k a^\dagger_k(t) a_k(t)) \label{eq:dens}
\end{eqnarray}
where $Z = {\rm Tr}( \exp(-\sum_k \lambda_k a^\dagger_k(t) a_k(t)))$. The trace here is taken on the complete 
Fock space while $a^\dagger_k(t)$ corresponds to creation operator of the canonical states $\varphi_k(t)$. In the filling approximation, 
the density operator corresponds to a statistical density and the information on the system reduces to the knowledge of the one-body
density matrix $\rho = \sum_k | \varphi_k(t) \rangle n_k \langle \varphi_k(t) |$ where the occupation numbers are related to the coefficients through
$n_k  =  1/(1+ e^{\lambda_k})$. The evolution of $\hat D(t)$ is performed by generalizing the TDHF approach where the single-particle states evolve
according to the standard self-consistent equation of motion (Eq. (\ref{eq:wf_ev})) while the occupation numbers are kept fixed in time.  As far as we know, this 
is the only way to avoid possible mixing of deformation and pairing effects and this procedure will be taken below as the no-pairing reference.     

Similarly to the pairing case, for non doubly magic nuclei, the density $\hat D(t)$ mixes systems with different particle numbers
and similar treatment based on double projections is necessary to extract transfer probabilities. In appendix \ref{app:proba}, 
some helpful formulas to perform projection on statistical densities of the form  (Eq. (\ref{eq:dens})) are given. An illustration of the 
decomposition of the initial state with a mean neutron number $\langle N \rangle = 26$ corresponding to the $^{46}$Ca is given in bottom
panel of Fig. \ref{fig:distN}. This figure illustrates that the width of the distribution is comparable to the BCS case (top panel) with the difference
that odd components are also present in the filling approximation. Probabilities obtained with the filling approximation will be labeled 
by $P_{xn}({\rm MF})$ while those with pairing will be labelled by $P_{xn}({\rm BCS})$.    

\begin{figure}[htbp] 
	\centering\includegraphics[width=\linewidth]{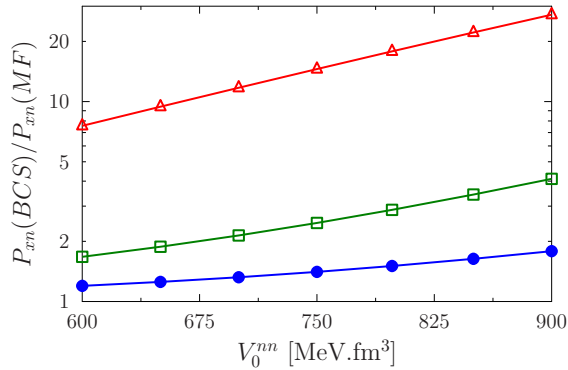}
  	\caption{Ratio of probabilities $P_{xn}({\rm BCS})/P_{xn}({\rm MF})$ as a function of 
the pairing strength interaction $V^{nn}_0$ for the reaction $^{40}$Ca+$^{46}$Ca at $E_{\rm c.m.} = 43.7$ MeV. $P_{1n}({\rm BCS})/P_{1n}({\rm MF})$ (filled circles) 
and $P_{2n}({\rm BCS})/P_{2n}({\rm MF})$
 calculated by neglecting (open squares) or not (open triangles) 
 the anomalous density components are shown.} 
 \label{fig:ratio} 
\end{figure}

As a first illustration of the enhancement of pair transfer probabilities when pairing is introduced, we have extracted 
systematically the ratios between probabilities with and without pairing as the pairing interaction strength $V^{nn}_0$ is varied in the mixed interaction
for the reaction $^{40}$Ca+$^{46}$Ca at $E_{\rm c.m.} = 43.7$ MeV. These ratios are shown in 
Fig. \ref{fig:ratio} as a function of $V^{nn}_0$. When pairing is accounted for, the two nucleons probabilities have been computed using either 
non-zero components of the anomalous density (open triangles) or neglecting them (open squares). While the former case corresponds to the appropriate 
treatment of pairing effects, the latter case can be regarded as a reference calculation where only the sequential transfer of the two neutrons is 
treated while taking properly the occupation number dispersion of single-particle states around the Fermi energy.
The pairing correlations strongly enhanced the two-particle transfer, by an order of magnitude around 
the physical value of the pairing strength (see table \ref{tab1:trans}). Note that the enhancement depends on the energy of the collision (see below). 
A smaller but non-zero effect is also seen in the one-particle transfer channel. The small increase in $P_{1n}$
stems from the increase of occupation number fragmentation as $V^{nn}_0$ increases.
The strong enhancement observed when the anomalous density is not neglected compared to the case where it is set to zero 
clearly shows that the increase is interpreted as the contribution from direct simultaneous processes. 
      
\section{Results and discussion}

In the present work, we have systematically investigated the effect of initial pairing correlations 
on the single- and multi-nucleon transfer by comparing the TDHF+BCS with frozen correlations 
to the mean-field dynamics with the filling approximation for collision between a $^{40}$Ca and different 
calcium isotopes below the Fusion barrier. In table \ref{tab:vb}, the fusion threshold energy $B_0$ deduced from 
the mean-field transport theories using the technique describes in ref. \cite{Sim08} are systematically reported for the different reactions considered here. When available, experimental fusion
barrier are also shown. It is clear from the table, that the introduction of pairing has a very weak influence on the barrier height. 
\begin{table}[!h]
\begin{center}
\begin{tabular}{|c|c|c|c|}
 \hline
     system & $B_0$(Exp.) & $B_0$ (Filling) [MeV] &  $B_0$(BCS) [MeV] \\
 \hline
 \hline
   $^{40}$Ca+$^{40}$Ca & 53.6 & 53.090  & 53.090 \\
    $^{40}$Ca+$^{42}$Ca&  & 52.735  & 52.735 \\
    $^{40}$Ca+$^{44}$Ca & 51.8 & 52.343  & 52.332 \\
    $^{40}$Ca+$^{46}$Ca&  & 52.069  & 52.049 \\
    $^{40}$Ca+$^{48}$Ca& 51.8 & 51.935  & 51.935 \\
    $^{40}$Ca+$^{50}$Ca&  & 51.200  & 51.247 \\
 \hline
 \hline
\end{tabular}
\end{center} 
\caption{Fusion barrier $B_0$ (in MeV) for the reaction $^{40}$Ca+$^{4x}$Ca.
Experimental barrier are taken from the systematic \cite{Siw04}, theoretical barrier are computed with a precision of 0.005 MeV. }
\label{tab:vb}
\end{table}

\subsection{Systematic study of two-particle transfer versus one-particle transfer}

\begin{figure}[htbp] 
	\centering\includegraphics[width=\linewidth]{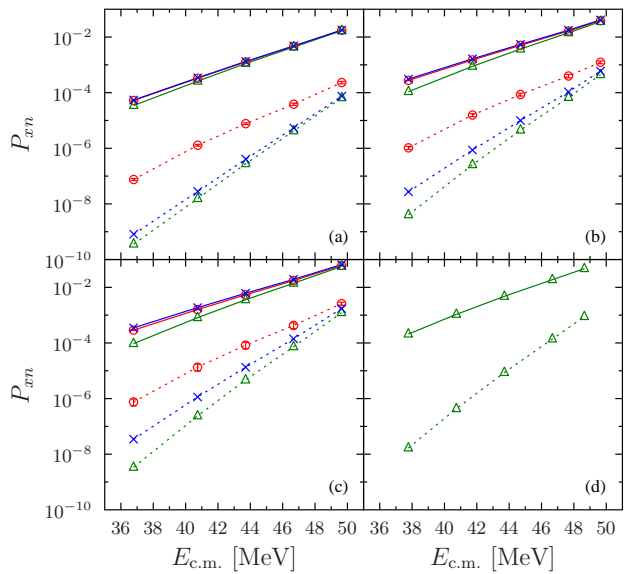}
  	\caption{One (dashed line) and two-particle (solid line) transfer probabilities as a function of center of mass energy 
	for the reactions (a) $^{40}$Ca+$^{42}$Ca, (b) $^{40}$Ca+$^{44}$Ca, (c) $^{40}$Ca+$^{46}$Ca and (d) $^{40}$Ca+$^{48}$Ca.
	The TDHF+BCS results obtained by neglecting (crosses) or not (open circles) the anomalous density contribution are 
	systematically compared with the mean-field case (open triangles). Note that for panel (d), both nuclei are closed shell nuclei 
	and pairing correlations vanishes. Accordingly, only the mean-field result is shown.
	} 
	\label{fig:ca4x} 
\end{figure}

In Figure \ref{fig:ca4x}, one- and two-particle transfer probabilities obtained for the collision between calcium isotopes
are  displayed as a function of center of mass energy for the TDHF+BCS case and no-pairing case.    
In all cases, when one of the collision partner presents pairing, the two-particle transfer probabilities are significantly enhanced. Conjointly, the 
one-particle transfer is also increased but to a less extend. This implies that the mean number of particles exchanged is also influenced by the pairing correlations
due to the sum-rule (Eq. (\ref{eq:sumrule})). Comparing the TDHF results where the effect of $\kappa$ is included (direct+sequential process) to the case where 
it is neglected (sequential only), several conclusions can be drawn. First, the one-particle probability is almost unchanged. 
Therefore, the enhancement in $P_{1n}$ observed in BCS theory 
compared to the pure mean-field case is a direct consequence of the specific fragmentation of occupation numbers due to pairing
that reduces Pauli blocking effect during the transfer process and is unaffected by the simultaneous component. 
A second important conclusion is that the main source of enhancement 
observed in $P_{2n}$ is coming from the initial two-body correlations themselves that lead to direct process during the collision. This confirms 
the observation made in Fig. \ref{fig:ratio}. 
      
\subsection{Correlations between two-particle transfer and pairing gap}

To further quantify the influence of pairing correlations on the enhancement of two-particle transfer and possible dependence with center of mass energy, 
the ratio $P_{2n}(BCS)/P_{2n}(MF)$ is displayed as a function of the mass of the heaviest nucleus participating to the collision and for two different fixed 
center of mass energies below the Coulomb barrier. For comparison, the neutron mean gap,
\begin{eqnarray}
\Delta_{BCS} = \frac{\sum_{k>0} \kappa_k \Delta_k}{\sum_{k>0} \kappa_k},
\end{eqnarray}
obtained for this nucleus is also shown in the top panel.
\begin{figure}[htbp] 
	\centering\includegraphics[width=\linewidth]{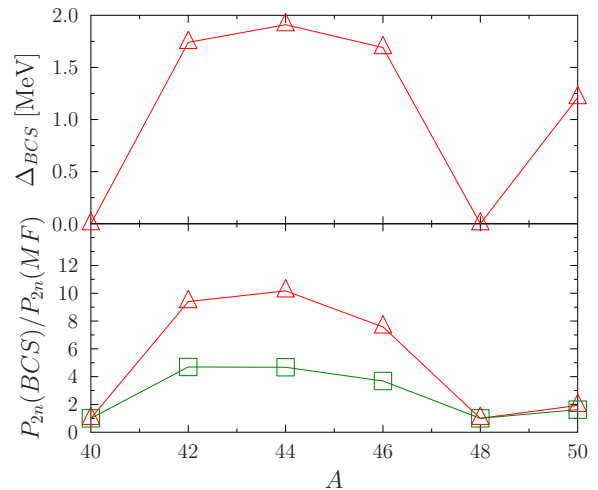}
  	\caption{Top: Mean neutron pairing gap obtained with BCS theory for the mixed interaction as a function
	of mass along the isotopic chain. Bottom: Ratio of the two-particle transfer probability obtained 
	with and without pairing at fixed center of mass energy below the Coulomb barrier reported in table 
	\ref{tab:vb}. Open triangles and open squares correspond to 4 MeV and 6 MeV below the Coulomb barrier respectively. } 
	\label{fig:ca4xbell} 
\end{figure}   
\begin{figure}[htbp] 
	\centering\includegraphics[width=\linewidth]{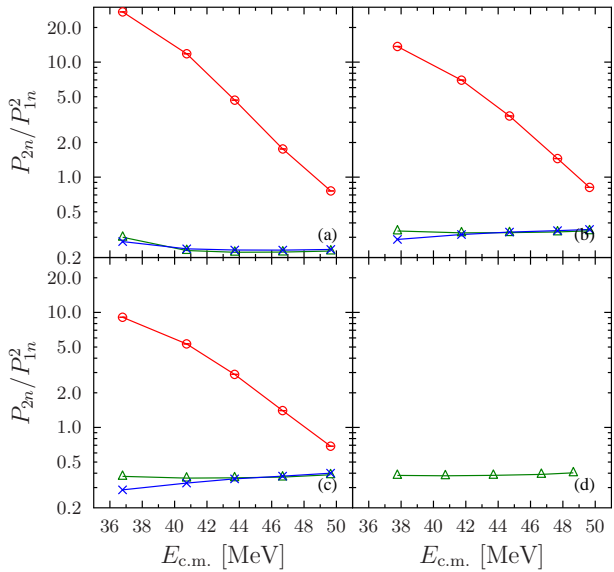}
  	\caption{Ratio $P_{2n}/(P_{1n})^2$ as a function of beam energies. The different panel correspond to different reactions:
	 (a) $^{40}$Ca+$^{42}$Ca, (b) $^{40}$Ca+$^{44}$Ca, (c) $^{40}$Ca+$^{46}$Ca and (d) $^{40}$Ca+$^{48}$Ca.
	 The TDHF+BCS results obtained by neglecting (crosses) or not (open circles) the anomalous density contribution are 
	systematically compared with the mean-field case (open triangles).
	  } 
	\label{fig:ca4xratios} 
\end{figure}   
Similarly to the pairing gap, this ratio has a typical bell shape that drops down to one in magic nuclei. This confirms that the enhancement 
of pair transfer is directly proportional to the initial pairing correlations (see for instance discussion in ref. \cite{Bri05}).

\subsection{Relationship between $P_{2n}$ and $P_{1n}^2$}
Experimentally, the no-pairing limit that would be a reference for a given reaction, cannot be measured. It is therefore important 
to compare quantities that could be measure simultaneously. Usually,  the two-particle transfer $P_{2n}$ is compared to $(P_{1n})^2$, where 
the latter quantity is considered as the probability for a completely sequential transfer \cite{Cor09,Cor11,Oer01}. Such a comparison has the 
advantage that both quantities contain all possible effects that might influence the transfer of particles as well as possible pollution from 
coming from experimental set-ups.  In figure \ref{fig:ca4xratios}, this ratio is presented for different theories considered here.   

This figure gives interesting insight in the two-particle transfer. First, both mean-field and TDHF+BCS where only the fragmentation 
of single-particle state  
is accounted for while $C_{12} = 0$, lead to almost identical ratios. This aspect was not clear from Fig. \ref{fig:ca4x} where different 
fragmentations obtained with the filling approximation and from BCS with $C_{12}=0$ lead to differences for both $P_{1n}$ and $P_{2n}$. 
The mean-field theory or equivalently the BCS where initial correlations are neglected could be considered as a way to mimic 
independent transfer of the two-particles.

It turns out that simple combinatorial arguments can be used to understand analytically the sequential limit.
Let us denote by $p$ the average probability to transfer one particle from the $^{4x}$Ca to $^{40}$Ca. Here 
"average" means that we disregard the fact the the probability depends on the initial and final single-particle states. It turns 
out that the total probability to transfer 1, 2, ...,  k nucleons
during the reaction  $^{4x}$Ca+$^{40}$Ca for $x>2$ in the MF 
approximation is consistent with:
\begin{eqnarray}
P_{1n} & = & \Omega_{1n} p (1-p)^{N_v-1} \nonumber \\
P_{2n} & = &  \Omega_{2n} p^{2}  (1-p)^{N_v-2}  \nonumber \\
\cdots \nonumber \\
P_{kn} & = &  \Omega_{kn} p^{k}  (1-p)^{N_v-k}  \label{eq:pkn}
\end{eqnarray}
where $N_v=x$ is the number of valence nucleons (with the constraint $k < N_v$) 
in the emitter with respect to the inert core of $^{40}$Ca, while $\Omega_{kn}$
is a purely combinatorial factor that depends on the number of nucleons in the valence shell and on the number of available single-particle states 
in the $f^{7/2}$ empty shell of the receiver nucleus ($N_f=8$). $\Omega_{kn}$ simply counts the number of possibilities to select $k$ particles among 
$N_v$ times the number of ways to put them in the $f^{7/2}$ shell, i.e.
\begin{eqnarray}
 \Omega_{kn}  &=& \frac{N_v!} {k! (N_v-k)! } \times N_f (N_f -1) \cdots (N_f - k + 1). \nonumber 
\end{eqnarray}
Accordingly, one can anticipate that 
\begin{eqnarray}
\frac{P_{2n}}{(P_{1n})^2} &=& \frac{1}{2} \frac{(N_v -1)}{N_v}  \frac{(N_f -1)}{N_f} \times \frac{1}{(1-p)^{N_v}} \nonumber \\
& \simeq &  \frac{1}{2} \frac{(N_v -1)}{N_v}  \frac{(N_f -1)}{N_f} \label{eq:factor}
\end{eqnarray}
where the last approximation holds  if $p \ll 1$.

\begin{figure}[htbp] 
	\centering\includegraphics[width=\linewidth]{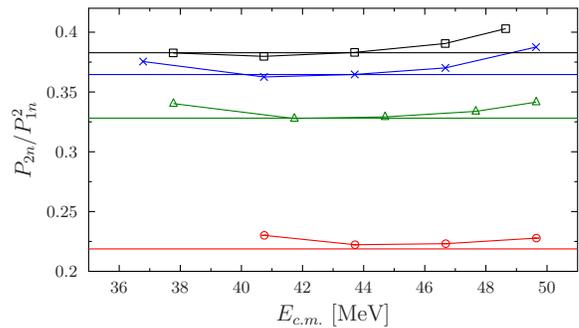}
  	\caption{Ratio $P_{2n}/(P_{1n})^2$ as a function of center of mass energy for the reaction $^{42}$Ca+$^{40}$Ca (open circles), 
	$^{44}$Ca+$^{40}$Ca (open triangles), $^{46}$Ca+$^{40}$Ca (cross) and $^{48}$Ca+$^{40}$Ca (open squares). The horizontal lines correspond 
	in each case to the value of left side of Eq. (\ref{eq:factor}) where $N_f=8$ while $N_v = x$ for $^{4x}$Ca+$^{40}$Ca reactions.} 
	\label{fig:p2p1bis} 
\end{figure}

This simple approximation turns out to work very well in the mean-field case (or equivalently in the pairing case when $\kappa$ is neglected). In figure 
\ref{fig:p2p1bis}, the quantity $P_{2n}/(P_{1n})^2$ is compared to the left side of Eq. (\ref{eq:factor}) for the different reactions considered here. We see that for a wide range of center of mass energy, mean-field results perfectly matches the relation (\ref{eq:factor}). 
The fact that such a simple description is adequate in mean-field theory is not trivial.
Indeed, in this theory, nucleons are quantal objects interacting first with two cores (the emitter and the receiver nucleus)
that are not fully inert and second with each other through the self-consistent mean-field. 
Last, the two transferred nuclei are fermions and are subject 
to the Pauli exclusion principle. This induces automatically correlations during the transfer. 
If a particle is already transferred to a certain single-particle level, this automatically 
forbid the other particles to be transferred to the same level. The latter effect is automatically 
included in the present theory and partially described through the factor $\Omega_{kn}$ in Eq. (\ref{eq:pkn}). 

Focussing now on the results accounting for initial correlations (open circles in Fig. \ref{fig:ca4xratios}), 
a strong, center of mass energy dependent, enhancement of the ratio is seen. The ratio increases 
significantly as the energy decreases from a value lower than 1 up to 20 in some cases.  
The present enhancement is at variance with the recent experimental observation in $^{40}$Ca+$^{96}$Zr where $P_{2n} \simeq 3 (P_{1n})^2$ has 
been observed almost independently of the center of mass energy \cite{Cor11}. It is worth mentioning however that the one- and two-particle transfer is anticipated to depend 
significantly on the structure properties, single-particle energies and spectroscopic factors, of the two collisions partners.

 In addition, here we are focussing 
on pairing correlation effect and paid a particular attention to not mix effects coming from static deformation in nuclei. Last, mean-field alone cannot grasp 
the physics of the quantum fluctuations in collective space. The inclusion of pairing partially cure this problem by increasing fluctuations of two-body 
observables. However, pairing alone does not contain all physical effects to treat this problem. This is clearly illustrated close to magicity where pairing vanishes.
In that case, TDHF dynamics is known to fail to reproduce transfer cross section. Recently, a stochastic mean-field approach was shown to properly 
describe quantal collective fluctuations especially in magic nuclei 
\cite{Ayi08,Ayi09,Was09,Yil11} and leads to realistic description of the nucleon exchange process. 
It would be interesting, in the near future to explore the possibility to 
combine stochastic methods with the present BCS approach. 

\section{Conclusion}

The TDHF+BCS theory with frozen correlations is used here to investigate the effect of pairing on one- and two-nucleon transfer 
below the Coulomb barrier. A method based on projection onto particle number is developed to properly extract transfer 
probabilities from theories that break the U(1) symmetry. In addition, a particular attention is paid to compare with a no-pairing 
limit free from possible effect of deformation. With this technique, the enhancement of two-particle transfer due to pairing correlations 
is studied qualitatively and quantitatively for reactions involving different calcium isotopes. It is shown, that when one of the collision partner has non-zero 
pairing, a strong enhancement of pair transfer is observed. This increase is directly proportional to the initial pairing correlations 
in the superfluid nucleus and turns out to strongly depend on the center of mass energy.     

\appendix
\section{Formulas for projection}
\label{app:proba}

In the present appendix, formulas useful for the numerical estimate of particle number projection 
are given for many-body quasi-particle states and density operators respectively given by Eqs. (\ref{eq:fullwf}) and (\ref{eq:dens}). 
\subsection{Particle number projection of density operators}

Starting from the density (\ref{eq:dens}), the probability to have $N$ particles in the subspace $B$ can be written as: 
\begin{eqnarray}
P_B(N) & = & \frac{1}{2\pi}\int_{0}^{2\pi}d\varphi e^{-i \varphi N}  {\rm Tr} \left( e^{i\varphi \hat{N}_B}  \hat D \right).
\end{eqnarray} 
The estimate of the trace can be made by writing the operator $\hat N_B$ in the canonical basis $\{\varphi_i \}$ associated 
to the density. 
Using the expression of $\hat N_B$ and the fact that the canonical basis forms a complete basis of the total single-particle space, it could be easily shown  
that:\begin{eqnarray}
P_B(N) & = & \frac{1}{2\pi}\int_{0}^{2\pi}d\varphi e^{-i \varphi N}  {\rm Tr} \left( e^{i\varphi  \sum_{ij} O^B_{ij}   a^\dagger_i a_j  }  \hat D \right), \nonumber \\
\end{eqnarray}
where 
\begin{eqnarray}
O^B_{ij} \equiv  \sum_\sigma \int d{\bf r} {\varphi_i}^*({\bf r},\sigma) {\varphi_j}({\bf r}, \sigma)   \Theta({\bf r}) = \langle i  | j \rangle_B.
\end{eqnarray}
Then, using formula (A.16) of ref. \cite{Bal85} leads to 
\begin{eqnarray}
{\rm Tr}\left( e^{i\varphi \hat{N}_B}  \hat D \right) &= & \frac1{z} \exp[{\rm Tr} \ln(1+e^{-i\varphi O^B}e^{-M})] \nonumber \\
&=& \frac1{z} {\rm det}(1+e^{-i\varphi O^B}e^{-M})
\end{eqnarray} 
where 
\begin{eqnarray}
\left(e^{-i\varphi O^B} \right)_{ij} =  F_{ij}(\varphi) = \delta_{ij} + \langle i | j \rangle_B (e^{i\varphi}-1),
\end{eqnarray}
while from formula (8.11) of ref. \cite{Bal85}, we have:
\begin{eqnarray}
(e^{-M})_{ij} &=& \delta_{ij} \frac{n_i}{1-n_i}
\end{eqnarray} 
and $z=\prod_{i}(1+\frac{n_i}{1-n_i} )$. Altogether, we obtain: 

\begin{eqnarray}
P_B(N) & = & \frac{1}{2\pi}\int_{0}^{2\pi}d\varphi e^{-i \varphi N} {\rm det} ( (1-n_j)\delta_{ij} + F_{ij}(\varphi) n_j ). \nonumber
\end{eqnarray}
Note that in the case where the statistical density identifies with a Slater determinant ($n_i = 0,1$), the formula given in ref. \cite{Sim10b}
is properly recovered.  Formulas for the double-projection technique can be derived using a similar technique. 

\subsection{Projection with quasi-particle states}

To perform projection of quasi-particle vacuum onto good particle number, we used the recently proposed Pfaffian method \cite{Rob09,Ave12,Ber12}. Since 
the Pfaffian technique has been largely discussed recently, here, only specific formulas useful in the present article are given.  Again, we first consider
the projection on the $B$ subspace as an illustration.  
We need to perform the overlap between the quasi-particle state (\ref{eq:fullwf}) and its gauge angle rotated counterpart: 
\begin{eqnarray}
| \Psi \rangle &=&\prod_{k>0} \left( u_k + v_ka^\dagger_k a^\dagger_{\overline{k}} \right)|-\rangle.  \nonumber \\
| \Psi_B (\varphi) \rangle &=& \prod_{k>0} (u_k+v_k b^\dagger_k (\varphi) b^\dagger_{\overline{k} }(\varphi)) |-\rangle , \nonumber
\end{eqnarray}
where 
\begin{eqnarray}
b^\dagger_i(\varphi) &=& \sum_\sigma \int d{\bf r} e^{i\varphi\Theta({\bf r})} \varphi_i ({\bf r},\sigma)\Psi^\dagger_\sigma({\bf r}) , \\
&=& \sum_j F_{ij}(\varphi) a^\dagger_j .
\end{eqnarray} 
The matrix ${\bf F}$ plays the role of the matrix ${\bf R}$ in ref. \cite{Ber12} and the overlap between the non-rotated and rotated state 
are given by Eq. (5) of this reference. In the present case, we obtain:
\begin{eqnarray}
\langle \Psi_0 | \Psi_B(\varphi) \rangle = \frac{(-1)^n}{\prod_{\alpha}^{n}v^2_{\alpha}} \rm{pf} \left[
\begin{array} {cc}
{\cal K} & {\cal M} (\varphi) \\
-{\cal M}^t (\varphi)  &  -{\cal K}^* 
\end{array} 
\right],\nonumber
\label{eq:pfaf}
\end{eqnarray}
where ${\cal K}$ and ${\cal M}$ are matrix of size $2n\times 2n$  where $n$ is the number of single-particle states with $i>0$. These matrices can be decomposed in $2\times2$ matrix blocks as:
\begin{eqnarray}
{\cal K}  & = & \left[ \begin{array} {cc}
{\bf  0} 			& \left[ \kappa_{i {\bar i}} \delta_{ij} \right]		\\
-\left[ \kappa_{i {\bar i}}\delta_{ij} \right]	& {\bf 0} 			
\end{array} 
\right],\nonumber
\end{eqnarray}
and 
\begin{eqnarray}
{\cal M} (\varphi)  & = & \left[ \begin{array} {cc}
\left[ v_i v_j F_{i j}(\varphi)  \right]		& \left[ v_i v_{\bar j} F_{i {\bar j}(\varphi) }   \right]	\\
\left[  v_{\bar i} v_j F_{{\bar i} j} (\varphi)  \right]	&	\left[ v_{\bar i} v_{\bar j} F_{{\bar i} {\bar j}}(\varphi)  \right]	
\end{array} 
\right],\nonumber
\end{eqnarray}
where matrix elements are directly indicated in each $n\times n$ block. 

For the double projection, the probability to find $N'$ particles in the space $B$ for a system of $N$ particles in the total space is given by
\begin{eqnarray}
P_B(N,N')=\frac{\langle N | {\hat P}_B(N') | N \rangle}{\langle N | N \rangle} = \frac{\langle \Psi | {\hat P}_B(N'){\hat P}(N) | \Psi \rangle}{\langle \Psi | {\hat P}(N) | \Psi \rangle} .
\end{eqnarray} 
Therefore,
we need to integrate with respect to two gauge angles.
\begin{eqnarray}
&&\langle \Psi | {\hat P}_B(N'){\hat P}(N) | \Psi \rangle = \nonumber \\
&& \frac1{4\pi^2} \int_{0}^{2\pi} d\varphi \int_{0}^{2\pi} d\varphi' e^{-i \varphi N - \varphi' N'} \langle \Psi | \Psi_B(\varphi,\varphi') \rangle.\nonumber
\end{eqnarray}
where  $\langle \Psi | \Psi_B(\varphi,\varphi') \rangle$ can be calculated using formula (\ref{eq:pfaf}) except that  $F_{ij}(\varphi)$ is now replaced by
$F_{ij}(\varphi,\varphi')= e^{i\varphi} F_{ij}(\varphi')$.

Numerically, the gauge integral are discretized using the Fomenko method \cite{Fom70} with 20 points. Note that during the time evolution, 
due to accumulated numerical errors, small violation of orthonormalization between single-particle states can occur, this might lead to 
large errors in the extracted transfer probabilities. To avoid this problem, a Gram-Schmidt orthonormalization algorithm is used prior to applying the Pfaffian formula.

\section*{Acknowledgment}  
We would like to thank G. Adamian, G. Antonenko, S. Kalandarov, G. Bertsch, D. Gambacurta, M. Grasso, V. Sargsyan, C. Simenel and K. Washiyama
for helpful discussions.

\end{document}